\documentclass[aps,showpacs,superscriptaddress,twocolumn]{revtex4}%
\usepackage{supertabular}
\usepackage{longtable}
\usepackage{amsfonts}
\usepackage{amsmath}
\usepackage{amssymb}
\usepackage{graphicx}%
\begin{document}
\title{Thermophysical properties of hydrogen-helium mixtures:
re-examination of the mixing rules via quantum molecular dynamics simulations}
\author{Cong Wang}
\affiliation{Institute of Applied Physics and Computational
Mathematics, P.O. Box 8009, Beijing 100088, People's Republic of
China} \affiliation{Center for Applied Physics and Technology,
Peking University, Beijing 100871, People's Republic of China}
\author{Xian-Tu He}
\affiliation{Institute of Applied Physics and Computational
Mathematics, P.O. Box 8009, Beijing 100088, People's Republic of
China} \affiliation{Center for Applied Physics and Technology,
Peking University, Beijing 100871, People's Republic of China}
\author{Ping Zhang}
\thanks{Corresponding author: zhang\underline{
}ping@iapcm.ac.cn}
\affiliation{Institute of Applied Physics
and Computational Mathematics, P.O. Box 8009, Beijing 100088,
People's Republic of China} \affiliation{Center for Applied
Physics and Technology, Peking University, Beijing 100871,
People's Republic of China}

\begin{abstract}
Thermophysical properties of hydrogen, helium, and hydrogen-helium
mixtures have been investigated in the warm dense matter regime at
electron number densities ranging from
$6.02\times10^{29}\sim2.41\times10^{30}$/m$^{3}$ and temperatures
from 4000 to 20000 K via quantum molecular dynamics simulations.
We focus on the dynamical properties such as the equation of
states, diffusion coefficients, and viscosity. Mixing rules
(density matching, pressure matching, and binary ionic mixing
rules) have been validated by checking composite properties of
pure species against that of the fully interacting mixture derived
from QMD simulations. These mixing rules reproduce pressures
within 10\% accuracy, while it is 75 \% and 50 \% for the
diffusion and viscosity, respectively. Binary ionic mixing rule
moves the results into better agreement. Predictions from one
component plasma model are also provided and discussed.

\end{abstract}


\maketitle

\setcounter{MaxMatrixCols}{10}

\setcounter{MaxMatrixCols}{10}

\bigskip

\section{Introduction}

The study of the equation of state (EOS), transport properties, and
mixing rules of hydrogen (H) and helium (He) under extreme condition
of high pressure and temperature is not only of fundamental interest
but also of essential practical applications for astrophysics
\cite{Hubbard:1981}. For instance, giant planets such as Jupiter and
Saturn require accurate EOS as the basic input into the respective
interior models in order to solve hydrostatic equation and
investigate the solubility of the rocky core
\cite{Wilson:2012:a,Wilson:2012:b}. On the other side, the evolution
of stars and the design of thermal protection system is assisted by
high precision transport coefficients of H-He mixtures at high
pressure \cite{Bruno:2010}. In addition, the viscosity and mutual
diffusion coefficients are also important input properties for
hydrodynamic simulations in modelling the stability of the hot
spot-fuel interfaces and the degree of fuel contamination in inertial confinement fusion (ICF)
\cite{Atzeni:2004,Lindl:1998}.

Since direct experimental access such as shock wave experiments is
limited in the Mbar regime \cite{Knudson:2009}, the states deep in
the interior of Jupiter ($\sim45$ Mbar) and Saturn ($\sim10$ Mbar)
\cite{Lorenzen:2009} can not be duplicated in the laboratory. As a
consequence, theoretical modelling provides most of the insight into
the internal structure of Giant planets. The EOS of H-He mixtures
have been treated by a linear mixing (LM) of the individual EOS via
fluid perturbation theory \cite{Stevenson:1975} and Monte Carlo
simulations \cite{Hubbard:1985}. Recently, several attempts have
been made to calculate EOS of H-He mixtures by means of quantum
molecular dynamic (QMD) simulations. Klepeis \emph{et al.}
\cite{Klepeis:1991} applied local density approximation of density
functional theory (LDA-DFT) calculations for solid H-He mixtures,
implying demixing for Jupiter and Saturn at 15000 K for a He
fraction of $x=N_{He}/(N_{He}+N_{H})=0.07$. Vorberger \emph{et al.}
\cite{Vorberger:2007}, Lorenzen \emph{et al.} \cite{Lorenzen:2009},
and Militzer \cite{Militzer:2013} performed QMD simulations by using
generalized gradient expansion (GGA) instead of the LDA in order to
evaluate the accuracy of the LM approximation and study the demixing
of H-He at Mbar pressures. Pfaffenzeller \emph{et al.}
\cite{Pfaffenzeller:1995} introduced Car-Parrinello molecular
dynamics (CPMD) simulations to calculate the excess Gibbs free
energy of mixing at a lower temperature compared to that set in the
work of Lorenzen \emph{et al.} \cite{Lorenzen:2009} and Militzer
\cite{Militzer:2013}.

Considering the transport properties, QMD \cite{Wang:2011} and
orbital-free molecular dynamics (OFMD) \cite{Recoules:2009}
simulations have been introduced to study hydrogen and its isotropic
deuterium (D) and tritium (T). Self-diffusion coefficients in the
pure H system and mutual diffusion for D-T mixtures were determined
for temperatures $T=$1 to 10 eV and equivalent H mass densities 0.1
to 8.0 g/cm$^{3}$
\cite{Collins:1994,Kwon:1994,Collins:1995,clerouin:1997,Kwon:1995,clerouin:2001,Kress:2010}.
QMD and OFMD simulations of self-diffusion, mutual diffusion, and
viscosity have recently been performed on heavier elements (Fe, Au,
Be) \cite{Lambert:2006:a,Lambert:2006:b,Wang:2013} and on mixtures
of Li and H \cite{Horner:2009}.

The present work selects H-He mixture as a representative system and
examines some of the standard mixing rules with respect to the EOS
and transport properties (viscosity, self and mutual diffusion
coefficients) in the warm dense regime that covers standard extreme
condition as reached in the interiors of Jupiter and Saturn. The
thermophysical properties of the full mixture and the individual
species have been derived from QMD simulations, where the electrons
are quantum mechanically treated through finite-temperature (FT) DFT
and ions move classically. In the next section, we present the
formalism for QMD and for determining the static and transport
properties. Then, EOS, viscosity, and diffusion coefficients for
H-He mixtures are presented, and the QMD results are compared with
the results from reduced models and QMD based linear mixing models.
Finally, concluding remarks are given.

\section{FORMALISM}
In this section, a brief description of the fundamental formalism
employed to investigate H-He mixtures is introduced. The basic
quantum mechanical density functional theory forms the basis of our
simulations. The implementation of schemes in determining diffusion
and viscosity is discussed. Mixing rules that combine pure species
quantities to form composite properties is also presented.

\subsection{Quantum molecular dynamics}
QMD simulations have been performed for H-He mixtures by using
Vienna \emph{ab initio} Simulation Package (VASP)
\cite{Kresse1993,Kresse1996}. In these simulations, the electrons
are treated fully quantum mechanically by employing a plane-wave
FT-DFT description, where the electronic states follow the
Fermi-Dirac distribution. The ions move classically according to the
forces from the electron density and the ion-ion repulsion.
Simulations have been performed in the NVT (canonical) ensemble
where the number of particles $N$ and the volume are fixed. The
system was assumed to be in local thermodynamic equilibrium with the
electron and ion temperatures being equal ($T_{e}=T_{i}$). In these
calculations, the electronic temperature has been kept constant
according to Fermi-Dirac distribution, and ion temperature is
controlled by No\'{s}e thermostat \cite{Hunenberger2005}.

At each step during MD simulations, a set of electronic state
functions [$\Psi_{i,k}(r,t)$] for each \textbf{k}-point are
determined within Kohn-Sham construction by
\begin{equation}\label{equation_eigenfunction}
    H_{KS}\Psi_{i,k}(r,t)=\epsilon_{i,k}\Psi_{i,k}(r,t),
\end{equation}
with
\begin{equation}\label{equation_hamiltonian}
    H_{KS}=-\frac{1}{2}\nabla^{2}+V_{ext}+\int\frac{n(r')}{|r-r'|}dr'+v_{xc}(r),
\end{equation}
in which the four terms respectively represent the kinetic
contribution, the electron-ion interaction, the Hartree contribution
and the exchange-correlation term. The electronic density is
obtained by
\begin{equation}\label{equation_electronicdensity}
    n(r)=\sum_{i,k}f_{i,k}|\Psi_{i,k}(r,t)|^{2}.
\end{equation}
Then by applying the velocity Verlet algorithm, based on the force
from interactions between ions and electrons, a new set of positions
and velocities are obtained for ions.

All simulations are performed with 256 atoms and 128 atoms for pure
species of H and He, and as for the case of the H-He mixture, a
total number of 245 atoms (234 H atoms and 11 He atoms) for a mixing
ratio $x=\frac{2N_{He}}{N_{H}+2N_{He}}=8.59\%$ (corresponding to the
H-He immiscibility region determined in the work of Morales \emph{et
al.} \cite{Morales2009}) has been adopted, where a cubic cell of
length $L$ (volume $V=L^{3}$) is periodically repeated. The
simulated densities range from 1.0 to 4.0 g/cm$^{3}$ for pure H
system. As for pure He and H-He mixture, the size of the supercell
is chosen to be the same as that for pure H to secure a constant
electron number density (in the range
$6.02\times10^{29}\sim2.41\times10^{30}$/m$^{3}$). The temperature
from 4000 K to 20000 K has been selected to highlight the conditions
in the interiors of Jupiter and Saturn. The convergence of the
thermodynamic quantities plays an important role in the accuracy of
QMD simulations. In the present work, a plane-wave cutoff energy of
1200 eV is employed in all simulations so that the pressure is
converged within 2\%. We have also checked out the convergence with
respect to a systematic enlargement of the \textbf{k}-point set in
the representation of the Brillouin zone. In the molecular dynamic
simulations, only the $\Gamma$ point of Brillouin zone is included.
The dynamic simulation is lasted 20000 steps with time steps of 0.2
$\sim$ 0.7 fs according to different densities and temperatures. For
each pressure and temperature, the system is equilibrated within 0.5
$\sim$ 1 ps. The EOS data are obtained by averaging over the final 1
$\sim$ 3 ps molecular dynamic simulations.

\subsection{Transport properties}
The self-diffusion coefficient $D$ can either be calculated from
the trajectory by the mean-square displacement
\begin{equation}\label{D_R}
    D=\frac{1}{6t}\langle|R_{i}(t)-R_{i}(0)|^{2}\rangle,
\end{equation}
or by the velocity autocorrelation function
\begin{equation}\label{D_v}
    D=\frac{1}{3}\int_{0}^{\infty}\langle V_{i}(t)\cdot V_{i}(0)\rangle dt,
\end{equation}
where $R_{i}$ is the position and $V_{i}$ is the velocity of the
$i$th nucleus. Only in the long-time limit, these two formulas of
$D$ are formally equivalent. Sufficient lengths of the trajectories
have been generated to secure contributions from the velocity
autocorrelation function to the integral is zero, and the mean
mean-square displacement away from the origin consistently fits to a
straight line. The diffusion coefficient obtained from these two
approaches lie within 1 \% accuracy of each other. Here, we report
the results from velocity autocorrelation function.

We have also computed the mutual-diffusion coefficient
\begin{equation}\label{D_mu1}
    D_{\alpha\beta}=\lim_{t\rightarrow\infty}\overline{D_{\alpha\beta}}(t)
\end{equation}
from the autocorrelation function
\begin{equation}\label{D_mu2}
    \overline{D_{\alpha\beta}}(t)=\frac{Q}{3Nx_{\alpha}x_{\beta}}\int_{0}^{t}\langle A(0)A(t')\rangle dt'
\end{equation}
with
\begin{equation}\label{D_A}
A(t)=x_{\beta}\sum_{i=1}^{N_{\alpha}}V_{i}(t)-x_{\alpha}\sum_{j=1}^{N_{\beta}}V_{j}(t),
\end{equation}
where the concentration and particle number of species $\alpha$ are
denoted by $x_{\alpha}$ and $N_{\alpha}$, respectively, and the
total number of particles in the simulation box
$N=\sum_{\alpha}N_{\alpha}$. The quantity $Q$ is the thermodynamic
factor related to the second derivation of the Gibbs free energy
with respect to concentrations \cite{Hansen1986}. In the present
simulations, $Q$ value has been adopted equal to unity since studies
with Leonard-Jones and other model potentials have shown that for
dissimilar constituents the $Q$-factor departs from unity by about
10\% \cite{Schoen1984}.

The viscosity
\begin{equation}\label{eta_lim}
    \eta=\lim_{t\rightarrow\infty}\bar{\eta}(t),
\end{equation}
has been computed from the autocorrelation function of the
off-diagonal component of the stress tensor \cite{Allen1987}
\begin{equation}\label{eta}
    \bar{\eta}(t)=\frac{V}{k_{B}T}\int_{0}^{t}\langle P_{12}(0)P_{12}(t')\rangle
    dt'.
\end{equation}
The results are averaged from the five independent off-diagonal
components of the stress tensor $P_{xy}$, $P_{yz}$, $P_{zx}$,
$(P_{xx}-P_{yy})/2$, and $(P_{yy}-P_{zz})/2$.

Different from the self-diffusion coefficient, which involves
single-particle correlations and attains significant statistical
improvement from averaging over the particles, the viscosity
depends on the entire system and thus needs very long trajectories
so as to gain statistical accuracy. To shorten the length of the
trajectory, we use empirical fits \cite{Kress2011} to the
integrals of the autocorrelation functions. Thus, extrapolation of
the fits to $t\rightarrow\infty$ can more effectively determine
the basic dynamical properties. Both of the $D$ and $\bar{\eta}$
have been fit to the functional in the form of
$A[1-\exp(-t/\tau)]$, where $A$ and $\tau$ are free parameters.
Reasonable approximation to the viscosity can be produced from the
finite time fitting procedure, which also serves to damp the
long-time fluctuations.

The fractional statistical error in calculating a correlation
function $C$ for molecular-dynamics trajectories
\cite{Zwanzig1969} can be given by
\begin{equation}\label{error}
    \frac{\vartriangle C}{C}=\sqrt{\frac{2\tau}{T_{traj}}},
\end{equation}
where $\tau$ is the correlation time of the function, and
$T_{traj}$ is the length of the trajectory. In the present work,
we generally fitted over a time interval of [0, $4\tau-5\tau$].

\subsection{Mixing rules}

Here, we examine two representative mixing rules. The first,
termed density-matching rule (MRd) with the inspiration of a
two-species ideal gas. The second, termed pressure-matching rule
(MRp), which follows from two interacting immiscible fluids.

In the MRd, the volume of the individual species is set equal to
that of the mixture ($V_{H}=V_{He}=V_{H-He}$), and QMD simulations
are performed for H at a density of $N_{H}/V_{H-He}$ and He at
$N_{He}/V_{H-He}$ at a temperature $T$. Then, pressure predicted by
MRd is determined by simply adding the individual pressures from the
pure species H and He simulations. Other transport coefficients,
such as mutual diffusion and viscosity, follow the same prescription
and are summarized as
\begin{equation}\label{eq_mrd}
\begin{split}
  V_{H-He} =& V_{H} = V_{He}, \\
  P_{H-He}^{d} =& P_{H} + P_{He}, \\
  D_{H-He}^{d} =& D_{H} + D_{He}, \\
  \eta_{H-He}^{d} =& \eta_{H} + \eta_{He}.
\end{split}
\end{equation}
The superscript is used to denote values predicted from MRd. The
derived pressure based on density mixing rule generally follows from
the ideal noninteracting H and He gas in a volume $V_{H-He}$.

The MRp has a more complicated construction compared to MRd. MRp can
be characterized as the following prescription:
\begin{equation}\label{eq_mrp}
\begin{split}
  V_{H-He} =& V_{H} + V_{He}, \\
  P_{H-He}^{p} =& P_{H} = P_{He}, \\
  D_{H-He}^{p} =& \nu_{H}D_{H} + \nu_{He}D_{He}, \\
  \eta_{H-He}^{p} =& \nu_{H}\eta_{H} + \nu_{He}\eta_{He}.
\end{split}
\end{equation}
In this case, we have performed a series of QMD simulations on the
individual species H and He, where the volumes change under a
constraint $(V_{H-He} = V_{H} + V_{He})$ until the individual
pressures equal to each other $(P_{H} = P_{He})$. The total pressure
becomes the predicted value. Here, we use the excess or electronic
pressure $P_{e}$ to evaluate this MRp mixing rule. Composite
properties such as mutual diffusion and viscosity are evaluated by
combining the individual species results via volume fractions
$(\nu_{\alpha}=V_{\alpha}/V_{H-He})$.

Finally, we also derive properties of the mixture from a slightly
more complex mixing rule \cite{Bastea2005}, as so-called binary
ionic mixture (BIM):
\begin{equation}\label{eq_bim}
\sum_{i}\nu_{i}\frac{\gamma_{i}-\gamma_{m}}{\gamma_{i}+\frac{3}{2}\gamma_{m}}=0
\end{equation}
with $\gamma$ the predicted mutual-diffusion coefficient or the
viscosity. The subscript $m$ denotes the mixture and $i$ the pure
species.

\section{Results And Discussion}
In this section, the wealth of information derived from QMD
calculations are mainly presented through figures, and the general
trends of the EOS as well as transport coefficients are concentrated
in the text. It is, therefore, interesting to explore not only to
get insight into the interior physical properties of giant gas
planets but also to examine a series of mixing rules for hydrogen
and helium. Additionally, one can consider the influence of helium
on the EOS and transport coefficients of mixing.

\subsection{The equation of state}

High precision EOS data of hydrogen and helium are essential for
understanding the evolution of Jupiter and target implosion in ICF.
Experimentally, The EOS of hydrogen and helium in the fluid regime
have been studied through gas gun \cite{Nellis2006}, chemical
explosive \cite{Boriskov2005}, magnetic driven plate flyer
\cite{Knudson2004}, and high power laser
\cite{Collins1998,Boehly2004,Hicks2009}. Since these experiments
were limited by the conservation of mass, momentum, and energy, the
explored density of warm dense matter were limited within $5\sim6$
times of the initial density. Recently, a new technique combined
diamond anvil cell (DAC) and high intensity laser pulse has
successfully been proved to provide visible ways to generate shock
Huguniot data of hydrogen over a significantly broader
density-temperature regime than previous experiments
\cite{Loubeyre2012}. However, the density therein was still
restricted within 1 g/cm$^{3}$.

\begin{figure}[!ht]
\includegraphics[width=1.0\linewidth]{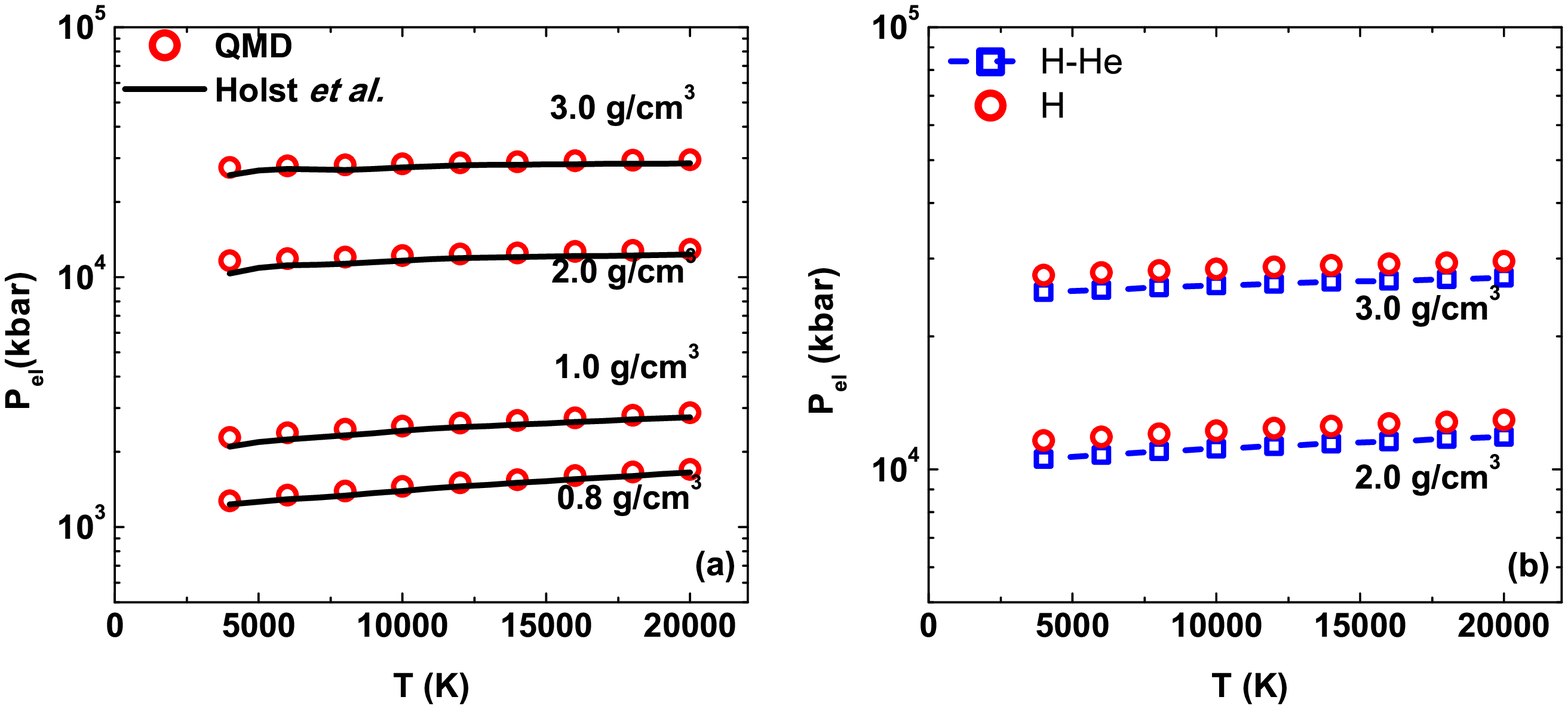}
\caption{(Color online) The present simulated QMD EOS have been
shown as a function of temperature (with error bar smaller than the
symbolic size). (a) Comparison with the results of Holst \emph{et
al.} \cite{Holst2008} for hydrogen; (b) The variation in EOS after
considering the mixing of helium. }\label{fig_eos1}
\end{figure}

In our simulations, wide range EOS for H, He, and H-He mixtures have
been determined according to QMD method. The EOS can be divided into
two parts That is, contributions from the noninteracting motion of
ions ($P_{i}$) and the electronic term ($P_{el}$),
\begin{equation}\label{eq_pressure}
    P_{total}=P_{i}+P_{el},
\end{equation}
where $P_{el}$ is calculated directly through DFT. In Fig.
\ref{fig_eos1} (a), we have compared our results of $P_{el}$ with
that of Holst \emph{et al.} \cite{Holst2008}, where the electronic
pressure is expressed as a smooth function in terms of density and
temperature, and the results agree with each other with a very
slight difference (accuracy within 5\%). In the simulated density
and temperature regime, we do not find any signs indicating a
liquid-liquid phase transition ($\frac{\partial P}{\partial
T}|_{V}<0$) or plasma phase transition ($\frac{\partial P}{\partial
V}|_{T}>0$), which are characterized by molecular dissociation and
ionization of electrons, respectively. With considering the mixing
of He into H, the electronic pressure is effectively reduced, as has
been shown in Fig. \ref{fig_eos1} (b).

\begin{figure}[!ht]
\includegraphics[width=1.0\linewidth]{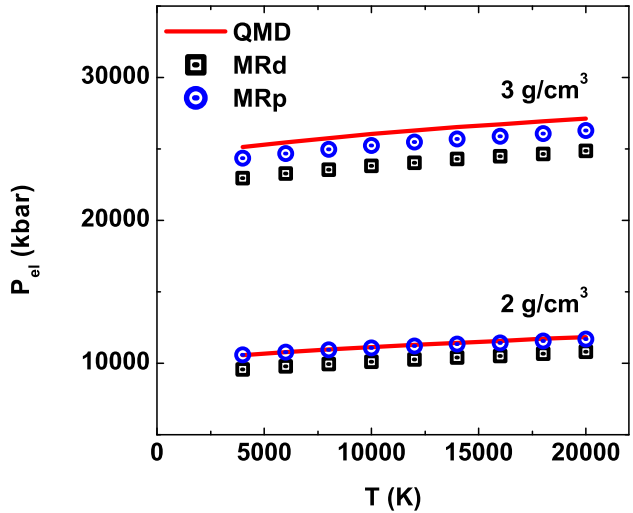}
\caption{(Color online) Comparison of electronic pressure between
direct QMD simulation results and those from mixing rules (error bar
being smaller than symbol size). QMD results are denoted by red
solid lines, while the MRd and MRp data are presented by open
squares and open circles, respectively.}\label{fig_eos2}
\end{figure}

MRd accounts for contribution from noninteracting H and He
subsystems in the volume of the mixtures. In MRd, the pressure
contributed from noninteracting ions is the same as that of the
mixture, but the electronic pressure is much lower due to the low
electronic density in the pure species simulations. It is indicated
that the electronic pressure is underestimated by MRd model at about
8\% $\sim$ 9\% (see Fig. \ref{fig_eos2}). For MRp model, we have
firstly performed a series of pure species simulations at a wider
density (temperature) regime compared to H-He mixtures. Then, the
simulated EOS data are fitted into smooth functions in terms of
density and temperature. Under the constraint of $V_{H-He} = V_{H} +
V_{He}$, we have predicted the electronic pressures $P_{H-He} =
P_{H} = P_{He}$ according to MRp model by solving pure species EOS
function at certain densities and temperatures, as shown in Fig.
\ref{fig_eos2}. It is indicated that the MRp model agrees better
with direct QMD simulations (accuracy within 3\%), the difference
mainly come from the ionic interactions between H and He species
after mixing.

\subsection{Diffusion and viscosity}

\begin{figure}[!ht]
\includegraphics[width=1.0\linewidth]{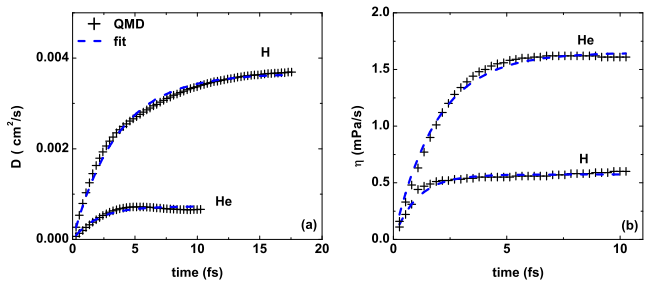}
\caption{(Color online) Transport properties are shown for H and He
at 12000 K with densities of 2 g/cm$^{3}$ and 8 g/cm$^{3}$,
respectively. (a) Self-diffusion coefficient; (b) Viscosity. The
direct QMD simulated results are presented by black crosses, while
the fitted results are denoted by blue dashed
lines.}\label{fig_dandeta}
\end{figure}

QMD simulations have been performed within the framework of FT-DFT
to benchmark the dynamic properties of H, He, and H-He mixture in
the WDM regime. Illustrations for the self-diffusion coefficients
and viscosity (for H and He at densities of 2.0 g/cm$^{3}$ and 8.0
g/cm$^{3}$, respectively) at a temperature 12000 K, as well as their
fits are shown in Fig. \ref{fig_dandeta}. The trajectory of the
present simulations lasts 4.0 $\sim$ 14.0 ps, and correlation times
between 1.0 and 15.0 fs. As a consequence, the computational error
for the viscosity lies within 10\%. After accounting for the fitting
error and extrapolation to infinite time, a total uncertainty of
$\sim$ 20\% can be estimated. The uncertainty in the self-diffusion
coefficients is smaller than 1\%, due to the additional
$\frac{1}{\sqrt{N}}$ advantage given by particle average.

\begin{figure}[!ht]
\includegraphics[width=1.0\linewidth]{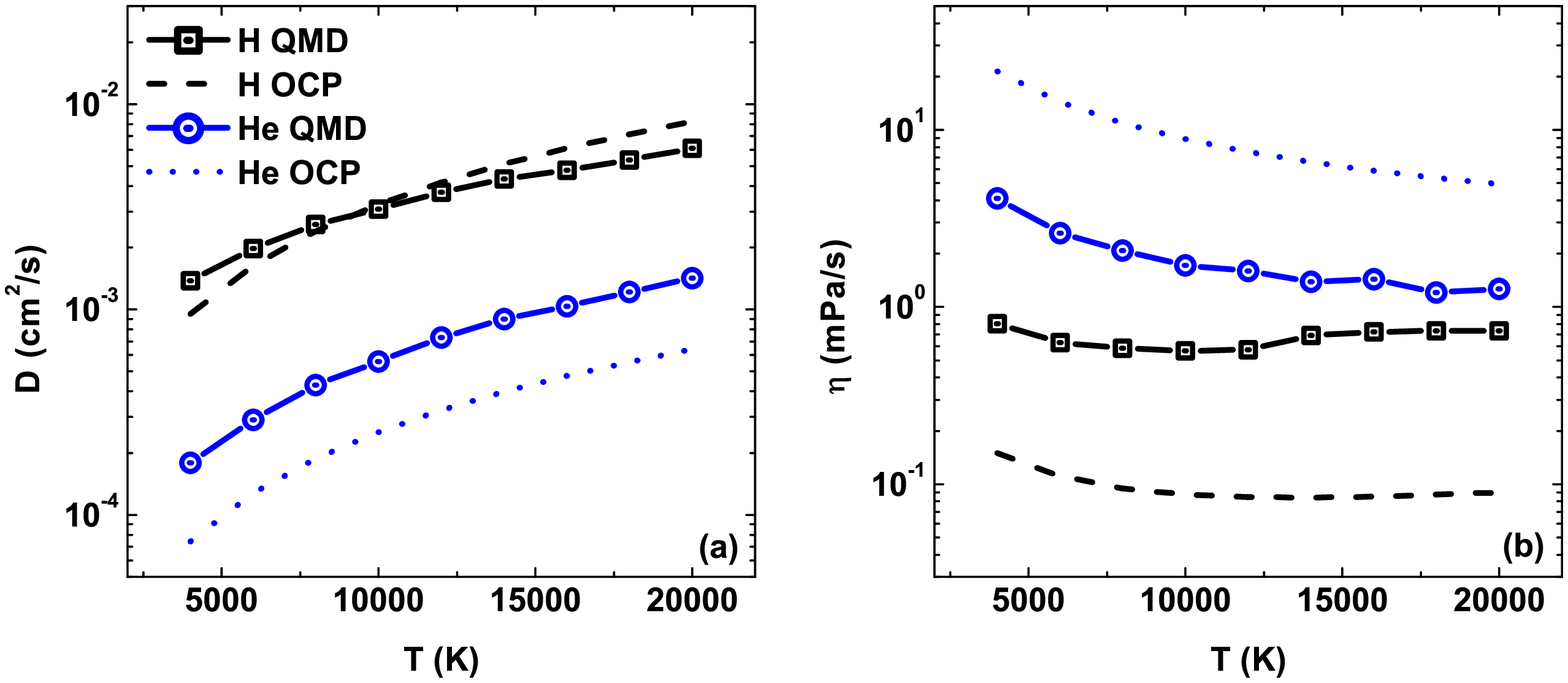}
\caption{(Color online) Comparison between QMD simulations and OCP
model (error bars are smaller than symbol size). (a) Self-diffusion
coefficient; (b) Viscosity. The sampled density for hydrogen is 2
g/cm$^{3}$. For helium, the density is 8 g/cm$^{3}$. The electron
number density for pure species (H and He) is $1.204 \times
10^{30}$/m$^{3}$. }\label{fig_qmd_ocp}
\end{figure}

Dynamic properties of WDM are generally governed by two
dimensionless quantities, namely, ionic coupling ($\Gamma$) and
electronic degenerate ($\theta$) parameter. The former one is
defined by the ratio of the potential to kinetic energy
$\Gamma=\frac{(Ze)^{2}}{ak_{B}T}$, with $Z$ the ionic charge, and
$a=(3/4\pi n_{i})^{1/3}$ the ion-sphere radius ($n_{i}$ is the
number density). The latter one $\theta=T/T_{F}$, where $T_{F}$ is
Fermi temperature. It has been reported that dynamic properties such
as diffusion coefficients and viscosity can be represented purely in
terms of ionic coupling parameter $\Gamma$ according to molecular
dynamics or Monte Carlo simulations based on one component plasma
(OCP) model
\cite{Bastea2005,Lambert2007,Bernu1978,Donko1998,Daligault2006,Hansen1975},
where ions move classically in a neutralizing background of
electrons. For instance, Hansen \emph{et al.} \cite{Hansen1975}
introduced a memory function to analyze the velocity autocorrelation
function, and obtain the diffusion coefficient in terms of
$D=2.95\Gamma^{-1.34}\omega_{p}a^{2}$ with the plasma frequency
$\omega_{p}=(4\pi n_{i}/M)^{1/2}Ze$. Based on classical molecular
dynamic simulations, Bastea \cite{Bastea2005} has fitted the
viscosity into the following form
\begin{equation}\label{eq_vis_ocp}
\eta=(A\Gamma^{-2}+B\Gamma^{-s}+C\Gamma)n_{i}M\omega_{p}a^{2},
\end{equation}
with $s = 0.878$, $A = 0.482$, $B = 0.629$, and $C = 0.00188$.

Since OCP model is restricted to a fully ionized plasma, we use
$Z$=1.0 (or 2.0) for hydrogen (or helium) to compute the
self-diffusion coefficient and viscosity. In Fig. \ref{fig_qmd_ocp}
(a), we show comparison between QMD and OCP model \cite{Hansen1975}
for hydrogen and helium at densities of 2 g/cm$^{3}$ and 8
g/cm$^{3}$. The general tendency for the self-diffusion coefficient
with respect to temperature is similar for QMD and OCP model,
however, the difference up to $\sim$60 \% is observed between the
two results. For the viscosity [Fig. \ref{fig_qmd_ocp} (b)], OCP
\cite{Bastea2005} predicts smaller (larger) values for hydrogen
(helium) compared to QMD simulations. The viscosity is governed by
interactions between particles and ionic motions, contribution from
the former one decrease with the increase of temperature, while, it
increases for the latter one. As a consequence, the viscosity may
have local minimum along temperature. For hydrogen, the local
minimum locates around 10000 K and 14000 K indicated by QMD and OCP
model, respectively. While in the case of helium, we do not observe
any signs for the local minimum in the simulated regime.

\begin{figure}[!ht]
\includegraphics[width=1.0\linewidth]{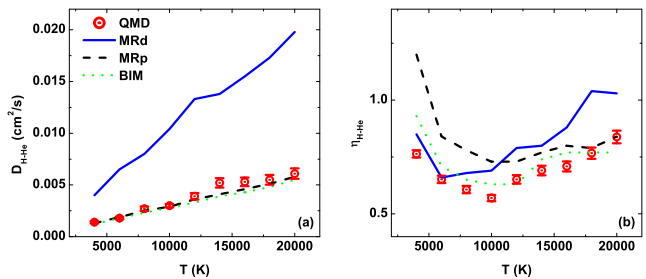}
\caption{(Color online) Comparison between QMD simulations and
mixing rules. (a) Mutual-diffusion coefficient; (b) Viscosity. The
sampled mixture has a electron number density of $1.204 \times
10^{30}$/m$^{3}$ and temperature from 4000 $\sim$ 20000
K.}\label{fig_mix}
\end{figure}

In Fig. \ref{fig_mix}, we have shown the mutual diffusion
coefficient $D_{H-He}$ and viscosity $\eta_{H-He}$ for H-He mixture
with an electron number density of $1.204 \times 10^{30}$/m$^{3}$,
and results from mixing models are also provided. The transport
coefficients predicted by MRd can be directly evaluated through Eq.
(\ref{eq_mrd}). For MRp model, we have firstly fitted the
self-diffusion coefficients and viscosity in terms of density and
temperature, after determining the volume for each species under the
constraint of $P_{H}=P_{He}$, the transport coefficients are then
obtained. In BIM model, we have used $\nu_{H}=91.41\%$ and
$\nu_{He}=8.59\%$, then, the transport coefficients are determined
by Eq. (\ref{eq_bim}). Here we would like to stress that in some
mixture studies based on average atom models \cite{{More1985}}, the
properties of pure species are derived from perturbed-atom models,
where boundary conditions are introduced from the surrounding medium
by treating a single atom within a cell. In the present work, the
dynamic properties of different mixing rules originate from QMD
calculations of the individual species. Despite divorced of the H-He
interactions, the pure species calculations still contain complex
intra-atomic interactions based on large samples of atoms.

The mutual diffusion coefficient of H-He mixture shows a linear
increase with respect of temperature, as indicated in Fig.
\ref{fig_mix}(a). The data from MRp and BIM models have a better
agreement with QMD simulations compared with that of the MRd model,
where ion densities are reduced and results in a larger diffusion
coefficient. The viscosity of H-He mixture has a more complex
behavior than pure species under extreme condition. As shown in Fig.
\ref{fig_mix} (b), MRd is valid at low temperature, while MRp works
at higher temperature. BIM rule moves the results into better
agreement with the H-He mixture, leaving within 30\% or better for
the simulated conditions.

\section{Conclusion}
In summary, we have performed systematic QMD simulations of H, He,
and H-He mixture in the warm dense regime for electron number
density ranging from
$6.02\times10^{29}\sim2.41\times10^{30}$/m$^{3}$ and for
temperatures from 4000 to 20000 K. The present study concentrated on
thermophysical properties such as the EOS, diffusion coefficient,
and viscosity, which are of crucial interest in astrophysics and
ICF. Various mixing rules have been introduced to predict dynamical
properties from QMD simulations of the pure species and compare with
direct calculations on the fully interacting mixture. We have shown
that MRd and MRp rules produce pressures within about 10 \% of the
H-He mixture, however, the mutual diffusion coefficients are as
different as 75 \% and it is 50 \% for the viscosity. BIM rule
generally gives better agreement with the mixture results. We have
also compared our QMD results with OCP model for the pure species.

\begin{acknowledgments}
This work was supported by NSFC under Grants No. 11275032, No.
11005012 and No. 51071032, by the National Basic Security Research
Program of China, and by the National High-Tech ICF Committee of
China.
\end{acknowledgments}

\end{document}